# A New Skill Based Robot Programming Language Using UML/P Statecharts

Ulrike Thomas, Gerd Hirzinger
Institute of Robotics and Mechatronics,
German Aerospace Center (DLR), Wessling

Bernhard Rumpe, Christoph Schulze
and Andreas Wortmann
Software Engineering, RWTH Aachen University

*Abstract*— This paper introduces the new robot programming language *LightRocks*(Light Weight Robot Coding for Skills), a domain specific language (DSL) for robot programming. The language offers three different level of abstraction for robot programming. On lowest level skills are coded by domain experts. On a more abstract level these skills are supposed to be combined by shop floor workers or technicians to define tasks. The language is designed to allow as much flexibility as necessary on the lowest level of abstraction and is kept as simple as possible with the more abstract layers. A Statechart like model is used to describe the different levels of detail. For this we apply the UML/P and the language workbench MontiCore. To this end we are able to generate code while hiding controller specific implementation details. In addition the development in *LightRocks* is supported by a generic graphical editor implemented as an Eclipse plugin.

## I. INTRODUCTION

The importance of flexible automated manufacturing grows continuously as products become increasingly individualized. Flexible assembly processes with robots are still hard to be coded which is due to many uncertainties caused among others by object tolerances, position uncertainties and tolerances from external and internal sensors. Possibilities to react on uncertainties demand for the application of sensors. As it is well known, the Light Weight Robot Arm (LWR) enables compliant motions and therewith helps to deal with uncertainties. Nevertheless programming the LWR is more complicate than programming known industrial robots, because not only positions but also stiffness and damping value need to be adjusted for each motion command. In addition, the complexity grows when more sensors are involved. Thus, only domain experts are able to program such compliant robot arms. At the moment the reusability of these programs depends on each individual expert. Tools allowing reusability and the composition of models at different level of detail are completely missing. Therefrom it exists a strong demand to develop such tools and to enable skill based robot programming for not well educated shop floor workers. To meet the requirements driven by easy to use software frameworks on the one hand and by employing many different sensors to achieve robustness on the other hand, we have developed a three layered architecture for robot programming based on Statecharts. Each skill, like grasping an object, mating two objects, screwing or else, can be coded by experts in a new skill based robot programming language. These skills can be selected and combined by non experts. In addition, an assembly planning system can be used to generate assembly sequences automatically. Each sequence is instantiated by a task recognizer. Each task can be modelled again by a Statechart, where skills are combined to robot tasks based on patterns. In particular for impedance controlled robots good programming tools are missing at the moment. With our new DSL *LightRocks* we achieve flexible and efficient programming of these robots, which is shown in two test cases. The robot programming language is based on UML/P Statecharts [1], [2]. The UML/P modelling language family is an implementation oriented variant of UML allowing code generation from several UML/P languages, e.g., Statecharts, class diagrams, OCL/P and others. Models of this language are generated into Java programs which are executable against a given robot programming interface using the language workbench MontiCore [3] and the UML/P code generation infrastructure [4]. The DSL itself is separated from the definition of the used robot interface. Therefore changes of the underlying robot interface only afflict the models not our DSL. To further improve its usability, the development of models in *LightRocks* is supported by a generic graphical editor framework in form of an Eclipse plugin. This paper is structured as follows: Section II introduces related work. Section III obtains the definitions of what we consider a task, skill and elemental action. In addition our DSL *LightRocks* is introduced. Section IV describes how MontiCore is used to generate code from *LightRocks* models. Afterwards Section V describes the case studies and Section VI concludes this paper.

## II. RELATED WORK

Since Mason specified robot compliant motions, it has become the goal of many researches to define compliant robot actions in an intuitive way [5]. The task frame formalism introduced by De Schutter and Bruyninckx [6] is an approach which became very popular. Meanwhile Hasegawa suggests skill primitives for each tiny motion [7]. Skill primitives have been constantly modified and improved so that complex robot task can be coded as nets of skill primitives [8]. The full task frame formalism has been applied and implemented for Cartesian force-feedback controlled parallel and serial robots [6], [9], [10]. Manipulation primitives net support-

This research has partly received funding from the European Union Seventh Framework Programme (FP7/2007-2013) under grant agreement n 287787, SMErobotics.

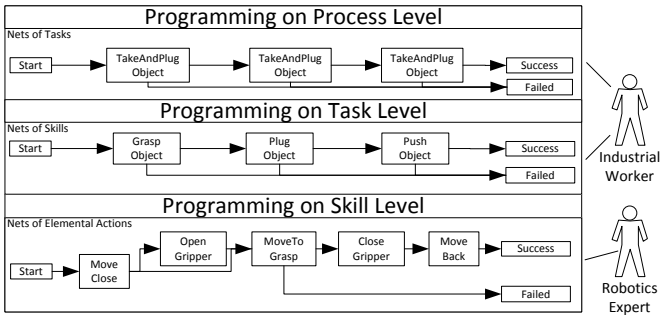

Fig. 1. The abstraction layers in our robot programming system.

ing an extended task frame formalism are implemented by Kroeger et al. [10]. In particular manipulation primitives are designed for force-feedback controlled robots. For the LWR these definitions need to be modified. Here, Statecharts are applied which are close to current modelling languages and suit better for code generation. Statecharts for robot programming have also been used by Branicky and Chhatpar [11]. They coded a typical peg-in-hole task by drawing Statechart diagrams and evaluated different search strategies which are again coded by Statecharts. Another intuitively to use interface for robot programming in object spaces is the iTASC-formalism suggested by Smits at al.[12]. The robot motions are specified in object space, where sensors are integrated in a uniform way. Statecharts are used on the controller level to allow reaction on different sensor values. In the meanwhile this approach has been improved by supporting inequality constraints and non-instantaneous task specifications [13] and its suitability is demonstrated on the PR2 by solving a co-manipulation task [14]. In another programming approach Kresse and Beetz [15] uses a symbolic description focusing on objects in the environment for robot programming and demonstrates its usefulness with a pan-cake making robot.

## III. DESCRIPTION OF THE *LightRocks* DSL

Fig. 1 illustrates our hierarchy for robot programming with different levels of abstraction. On the lowest level, instantiated actions are used, which contain primitive commands. These actions are combined to skill nets, similar to the description of skill primitive nets or manipulation primitive nets. For more details on this topic, we refer to [8], [10]. Based on these definitions and according to the requirements given by the LWR, we define the elements of *LightRocks*.

### A. The Interface to the LWR

The interface to the LWR consists of a set of controller parameters for each motion command to be send via TCP/IP to the control unit. The controller offers motions in joint space and in Cartesian space. The applied impedance controller is described in [16]. For motions in contact, we usually apply Cartesian motions. Our programming interface store object information about the entire robot workspace in form of a scene graph. A Cartesian motion command consists of following elements:

- the task frame $TF := \{T, Ref, Link\}$ with $T \in \mathbb{R}^{4 \times 4}$ and $Ref$ can be every object in the scene graph of the robotics environmental model.
- a goal frame $GF := \{T, Ref\}$, where the reference need not to be the same as for the task frame.
- a set of stiffness values $ST \in \mathbb{R}^6$ and damping values $DA \in \mathbb{R}^6$, according to the impedance controller described in [16].
- a stop condition: $\lambda \to \{true, false\}$ which maps sensor values to a boolean value. It is a conjunction or disjunction about measurable values.

In cases where the motion command is specified in joint space the task and goal pose are described by a set of seven joint variables and the sets for stiffness and damping values are needed for each joint respectively.

### B. Elements of LightRocks

As basic element of *LightRocks* we specify Elemental Actions (EA) . These are our tiniest actions and they map from the semantic point of view to a motion interface almost directly. Although an elemental action can be more than this, because further devices like tools or perception services can be addressed. An EA is defined as follows:

*Definition 3.1 (Elemental Action (EA)): An EA is a tuple $\langle Device, DeviceComands, \lambda, \mu \rangle$. It assigns a device to a certain command. The action should stop when the condition $\lambda$ triggers. The return values $\mu$ update our environmental model $EM$. The $Device$ is an element of any available physical control unit. The $DeviceCommand$ is adapted to the selected device.*

In our current implementation the $Device$ is of $\{Tool, Robot, PerceptionUnit\}$. In the case the device is a robot, the command has following elements $\langle flag, TF, GF, JV, ST, DA \rangle$. We use the flag to distinguish between motions in joint space and in Cartesian space. All the other elements are defined similar to the interface definition. Therewith, we are able to code EAs to be executed by an activated device which is either a robot, a tool or a sensor service. For example, we offer sensor services able to localize known objects. We call the recognition process and return the estimated pose of an object into our environmental model. In order to grasp the object, we need a second EA. It moves the robot into the approach pose. A third EA is necessary to grasp the object by applying Cartesian impedance control. With the Cartesian impedance controlled motion we are able to deal with uncertainties, which may arise from the pose estimation process. Hence complex skills, like recognizing and grasping objects can be coded as a net of EAs. We call such nets, which offer a service by our robotic system, skills. In the following we give a definition for a skill:

*Definition 3.2 (Skill): A skill is a capability of our robotic system. It is specified as a net of EAs: $Skill := \langle Nodes^*, Transitions^*, StartNodes^*, StopNodes^* \rangle$ with Nodes being a set of EAs and transitions are the connectors between them. At least one Node is within the StartNode set and the same yields for the StopNode set.*

Transitions have a pre-condition and a post-condition. Between this conditions the environmental model is updated by the return values of each Node. With theses Skills complex robot tasks can be coded. Usually a robotic task follows the order of grasping an object, transferring it to a certain pose and manipulating the object there. Then the robot should take the next object and so on. For each element of such sequence a skill can be used. In order to allow exception handling, we define a task also as a combinations of Skills:

*Definition 3.3 (Task):* A Task can be described as a net of Skills: $Task := \langle Skills^*, Transitions^*, StartSkills^*, StopSkills^* \rangle$. Each Task has at least one StartSkill and one StopSkill. The transitions are equal to those of Skills, with exactly the same behaviour.

The Skills are supposed to be coded by robot experts as net of the EAs where the Tasks can be implemented by combining theses Skills. For that, no detailed knowledge about the controller and the robot devices are necessary. The coding of skills suits well for graphical programming with easy to use tools. On a more abstract level the reordering of tasks can easily be done, if necessary. Thus, a shop floor worker might be able to program the system by stitching Skills or Tasks together.

### C. Describing the Elements of LightRocks with the UML/P

A declarative semantic might be used to describe the elements of *LightRocks*, here we have applied the UML/P, a formal semantics, for the specification of the DSL, refer to [17]. The UML/P modeling language family consists of six languages derived from standard UML [18]: class diagrams, Statecharts, sequence diagrams, object diagrams, Java code and the object constraint language (OCL), where Java is integrated into the UML/P to achieve a programmability. Statecharts [19] are finite automata used to model the behavior of a system. The similarity between nets of EAs and Statecharts allows to model Skills with Statecharts while reusing the existing UML/P code generation infrastructure [4]. UML/P Statecharts consists of *states* and *transitions*. States describe object or method states in object-oriented programming. Each state features a *name*, an *invariant*, an *entry action*, an *exit action* and an *activity*. States may further be composed from other states, allowing arbitrary state hierarchies. Transitions describe state changes from a source state to a target state. Each transition features a *stimulus*, an optional *action*, an optional *precondition* and an optional *postcondition*. The stimulus describes when this transition may activate and the action may be performed during transition from source state to target state. Fig. 2 illustrates the similarities to Skills or Tasks: both describe a system state and transitions with conditions restricting how the system may behave. Projecting Skills and Tasks on UML/P Statechart syntax allows to reuse much of the existing UML/P tooling [4]. Thus, the textual representation of *LightRocks* resembles Statechart syntax closely: Tasks, Skills and Elemental Actions are projected onto states, the directed arcs are mapped onto transitions and start and stop conditions are mapped onto the preconditions of entry actions and exit actions respectively. In order to apply the UML/P for each level of abstraction in our language, we need one single generic component, which is a superposition of the elements of *LightRocks*.

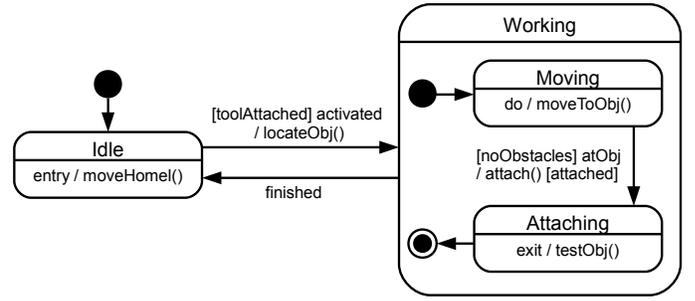

Fig. 2. A hierarchical Statechart illustrating the UML/P Statechart language elements.

### D. Generic Component

For the programmers point of view we have separated the elements in different levels of abstraction. For code generation it is more convenient to have one super component called Generic Action Component:

*Definition 3.4 (Generic Action Component):* A Generic Action Component is a tuple $(l, A, B, P, N, e)$, where:

- $l$ is a level identifier.
- $P$ is a set of parameters, which are defined by type, name identifier and an optional default value.
- $A$ is a set of start conditions, where the corresponding Action Component can be started if one $a \in A$ holds and this condition is connected.
- $B$ is a set of end conditions, where the corresponding Action Component is finished if one $b \in B$ holds and this condition is connected.
- $N = \{C, T\}$ consists of a set of child Action Components $C$ and a corresponding set of transitions $T$. Each transition is defined between end and start conditions of the children or the parent Action Component. In addition on each transition values can be assigned to the parameters of the target Action Component.
- $e \in Methods$ is a execution command, which describes the execution of this action with a concrete method call. Undefined if $C$ is not empty.

A generic action component can be considered as a kind of method declaration with additional conditions, describing in which situations the declared method can be called and in which situations this method is interrupted or finished. The declaration is combined with the methods definition directly, which is either expressed by other methods defined as child action components or a concrete method call defined in the related *domain interface*. In Fig. 3 an action component is exemplary visualized to illustrate the defined structure.

## IV. CODE GENERATION

Each Action Component is mapped to one Statechart, while the defined child components and conditions are expressed by states. For each UML/P Statechart an additional code segment can be defined, which is used to list the parameter declarations of the Action Component. To reuse or extend Action Components in different contexts this feature has to be established for UML/P Statecharts, too. Therefore additional stereotypes for states are introduced. These stereotypes indicate a reference to different Statecharts, instead of a concrete state definition inside the current model. Transitions are expressed by transitions and stereotypes identify the different layers. Based on this mapping Action Components are projected to UML/P Statecharts. To execute Action Components, code is generated from the corresponding UML/P Statechart. The generated code interacts with two different runtime environments: One implementing the interfaces described in the corresponding domain interface and one supporting the generated code to realize state hierarchies, transition functionality and import or extension of states. The first runtime environment is an adapter between the domain interface and the API of the robot, which shall be accessed by the Action Components, as illustrated in Fig. 4. For each *domain interface* different adapters can be implemented to support several robot APIs. The corresponding adapter called by the generated code can be switched easily, resulting in high re-usability of defined models. The approach to realize the second runtime environment is similar to the state pattern [20] It is illustrated in Fig. 5. Every state has a current state and a parent state. Transitions are defined as an own class, storing source and target state. For each model the corresponding states and transitions are generated, each inheriting from the defined base class in the state runtime environment. A generated state overwrites functions to define its child states, corresponding transitions, the entry action and entry condition, if given. Each imported state is an instance of a concrete state class generated from a model and each extended state is an instance of a concrete state class, which is not directly inherited from the base abstract state class, but from another concrete state class. The classes of the runtime environment are defined close to the state of the UML/P Statechart language that most parts can be translated one by one and the resulting templates are clearly arranged.

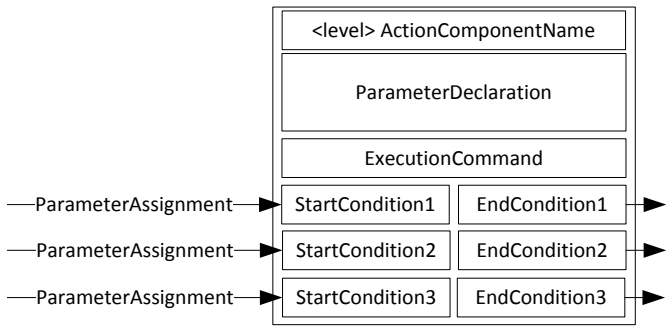

Fig. 3. Graphical representation of the generic Action Component definition.

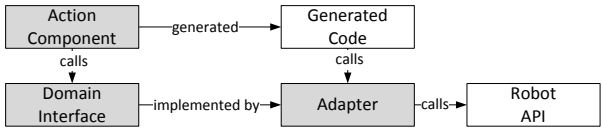

Fig. 4. Relationship between Action Component, domain interface and the adapter of the robot API.

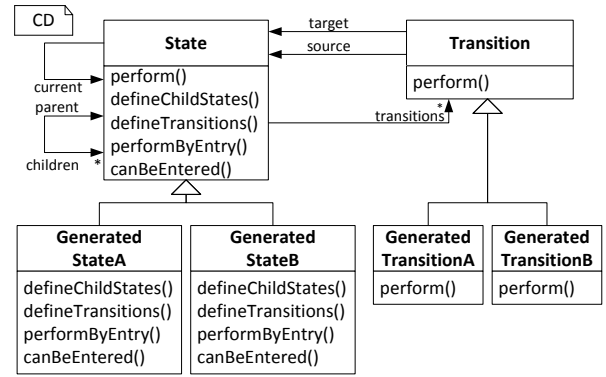

Fig. 5. Generated code and its state runtime environment.

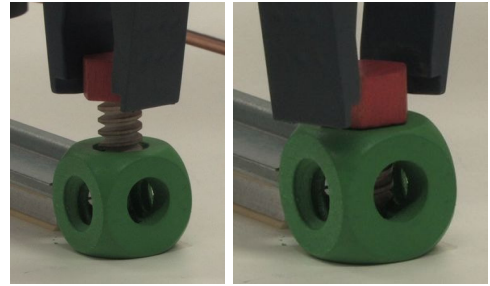

Fig. 6. The figure shows two different steps during the screwing Skill.

## V. CASE STUDIES

For evaluating the programming environment two test cases are chosen. The first test case is about screwing a wooden screw into a cube, see Fig.6. These objects are from wooden toys, but suit very well to demonstrate the capabilities of our new programming environment. The implementation on Task level and on Skill level are illustrated in Fig.7, Fig.8 respectively.

The screw is gripped after localizing, therefore the output of the perception service is updated into our environmental model. Then the robot moves the screw into the approach pose above the cube. The first part of the screwing skill starts by moving the object down the screw-axis until a torque about $0.32$ Nm has been exceeded. The gripper opens and the robot turns its hand back about $-180°$. Then it starts to screw again. This is repeated until a certain position in $z$-axis has been reached or the stop-condition triggers. If errors occur like the object is not gripped, the Statechart can react and the execution proceeds with another skill or finishes. The second case study is an assembly, where electrical parts have to be plugged on a top hat rail. A perception service is applied for the localization of objects. The perception units reads RGBD-data and matches geometric models in the acquired point cloud. Therefrom, we obtain the pose of the object and use the impedance controller to reduce uncertainties. Fig.9 shows the necessary task for achieving the assembly goal and Fig.10 depicts the Statechart of a single task which is a combination of Skills. Hence new tasks can be designed by reusing predefined skills. In this case programming such a task has been done by less than an hour. In addition, tasks can be reordered easily and moreover, models of skills can be reused for new skills by changing only as few elements as necessary. For instance, if the job floor worker will change the sequence of execution, only the tasks need to be reordered.

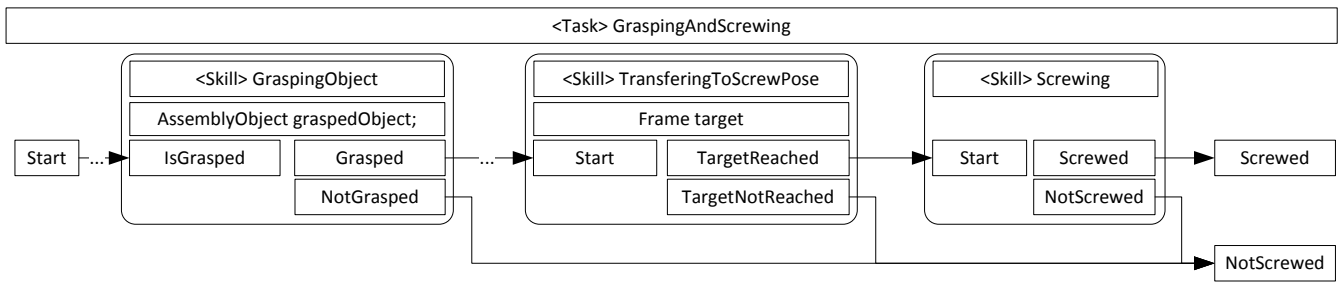

Fig. 7. The Task for grasping an object and screwing

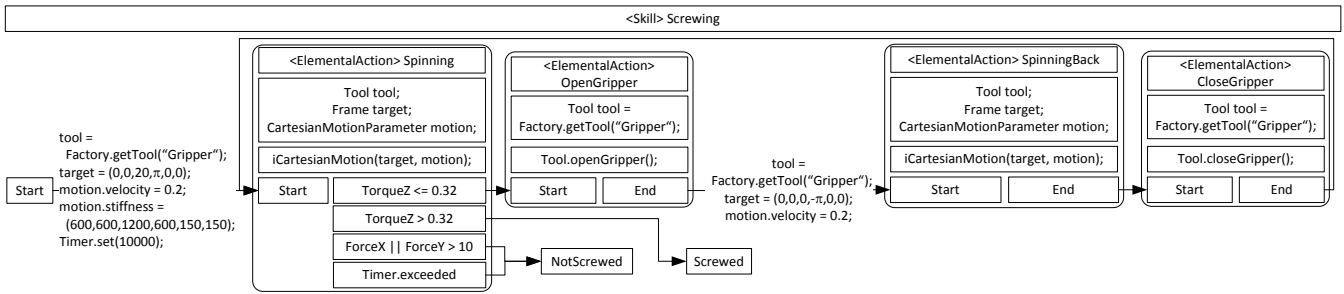

Fig. 8. This Statechart diagram shows the implementation of the skill for screwing

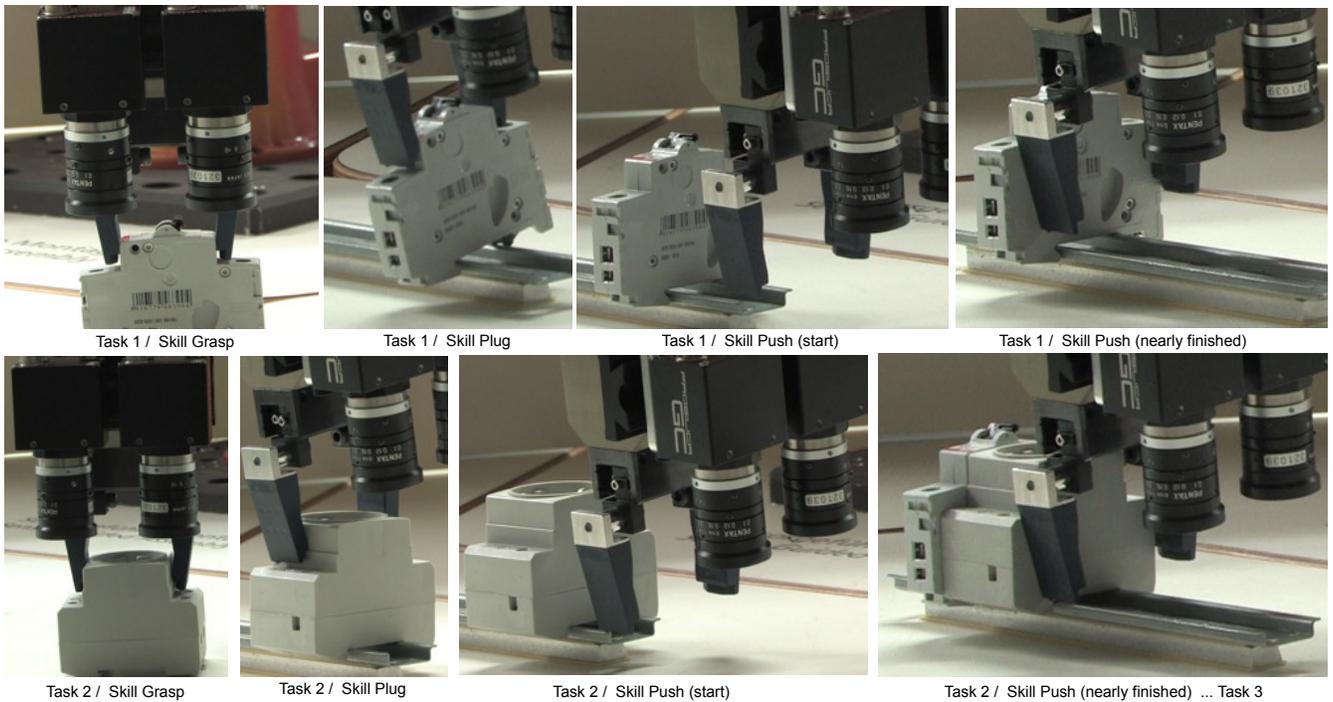

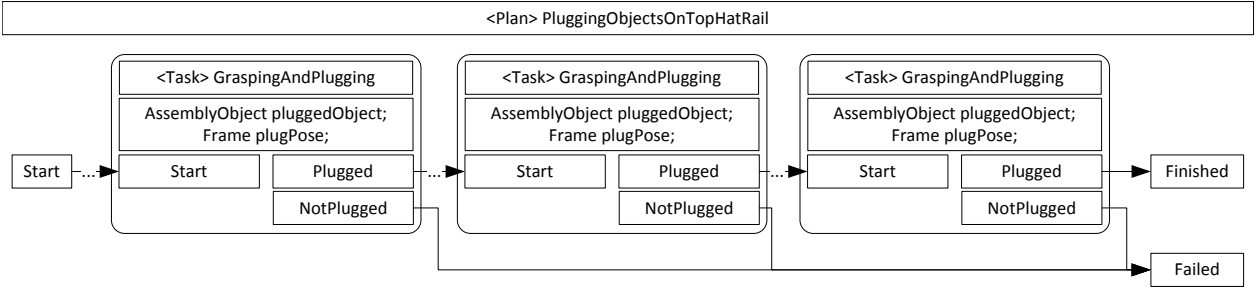

Fig. 9. The sequence for assembly is shown which containing three tasks. The task implementation is illustrated in the Fig.10. The images show the execution process for the first two parts.

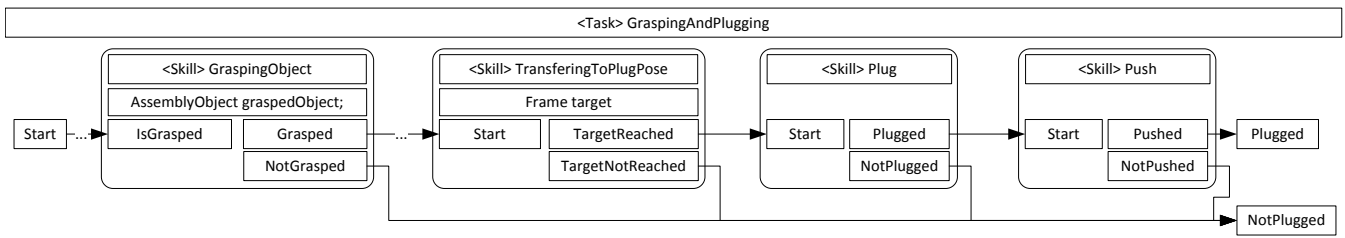

Fig. 10. A task for grasping and plugging an electrical object on the top hat rail is shown, which is embedded in the entire sequence.

## VI. CONCLUSIONS

The two test case scenario are solved with the LWR successfully by our new robot programming environment. The necessary elements of Tasks, Skills and Elemental Actions can be implemented efficiently using our new language *LightRocks*. It is demonstrated, that defining assembly processes and assembly tasks is quite intuitive, even for someone who is not very familiar with robotics. *LightRocks* allows reusability and intuitive programming at all supported levels of abstraction. The implementation of *LightRocks* as a MontiCore language further allows to reuse the models with generators for different target platforms. Our design of the generic action component and the usage of UML/P offers high flexibility for supporting other robots than the LWR. In cases where a Cartesian force-feedback controlled robot should be programmed, the necessary changes are small due to the generic action component and the UML/P. Instead of defining and implementing a complete new system the framework of MontiCore is employed intensely, with all its advantages and used for generating code for the given DSL. The designed DSL *LightRocks* and its supporting tools improve the development of applications for the LWR. In addition, it can be reused in many different contexts and for different robots. The DSL *LightRocks* can flexibly be extended to support additional sensors or robot motion specifications. In the future we plan to evaluate our system in real assembly lines, where non experts shall become able to program robots. In addition, we will improve the language of *LightRocks* for allowing more concurrency on Statechart level and compare our approach to others like sequential function charts defined in the IEC-61131.